\documentstyle[preprint,aps,epsfig]{revtex}
\tighten

\begin{document}

\title{Recoil growth: an efficient simulation method for multi-polymer
  systems}

\author{S. Consta$^1$, N.B. Wilding$^{2,3}$, D. Frenkel$^1$ and Z.
  Alexandrowicz$^{4,3}$}

\address{$^1$ FOM Institute of Atomic and Molecular Physics (AMOLF),  Kruislaan 407, 1098 SJ Amsterdam, The Netherlands.\\
  $^2$ Department of Physics and Astronomy, The University of Edinburgh,\\ Edinburgh EH9 3JZ, U.K.\\
  $^3$ Institut f\"{u}r Physik, Universit\"{a}t Mainz, Staudinger Weg 7,\\ D-55099 Mainz, Germany.\\
  $^4$ Department of Materials and Interfaces, Weizmann Institute of
  Science,\\ Rehovot 76100, Israel.  }

\date{\today}

\maketitle

\begin{abstract}
  
  We present a new Monte Carlo scheme for the efficient simulation of
  multi-polymer systems. The method permits chains to be inserted into
  the system using a biased growth technique. The growth proceeds via
  the use of a retractable feeler, which probes possible pathways ahead of
  the growing chain. By recoiling from traps and excessively dense
  regions, the growth process yields high success rates for both chain
  construction and acceptance. Extensive tests of the method using
  self-avoiding walks on a cubic lattice show that for long chains and
  at high densities it is considerably more efficient than
  Configurational Bias Monte Carlo, of which it may be considered a
  generalisation.

\end{abstract}

\section{Introduction}

Much current activity in polymer science focuses on systems of
mutually interacting polymer chains, such as semi-dilute and
concentrated solutions, melts, or chains tethered to a surface. Topics
of interest include chain conformations, phase behavior, and
relaxation dynamics with emphasis on entanglement, reptation and the
glassy state.  Besides the use of analytical methods, advances in the
theoretical understanding of such systems draw heavily on computer
simulation \cite{BOOK}.  Unfortunately, simulations of multi-chain
polymer systems are notoriously difficult and except for the lowest
densities, are largely restricted to unrealistically short chains
\cite{KREMER,LEONTIDIS}.

The principal difficulty is one of chain entanglement.  Large scale
conformational rearrangements of a chain are hindered by excluded
volume restrictions both with itself and with neighbouring chains.
This, of course, merely reflects the prevailing physical situation:
multi-chain dynamics are exceedingly slow on the time scales found in
simple molecular systems. Consequently, simulation methods such as
Molecular Dynamics (MD) which attempt to model the true physical
dynamics, suffer from relaxation times that increase rapidly with
chain length and density \cite{KREMER}. The same problem applies to
Monte Carlo (MC) methods employing random local motion of chain
segments, such as bond rotation algorithms or the bond fluctuation
model \cite{KREMER}.

In many situations however, one is not explicitly interested in the
intrinsic dynamics of the polymer system, but merely its static
equilibrium properties. Alternatively one may wish to study
relaxation, but lacks the means of efficiently preparing well
equilibrated starting configurations.  In such instances it is
expedient to forego local MD and MC algorithms in favour of {\em
  artificial} MC dynamics, which permit a much more efficient
exploration of configurational phase space. Unfortunately, most of the
efficient algorithms tailored for a single chain, such as the pivot
\cite{LAL}, cannot be applied in the multi-chain context. One
exception is the well known `slithering snake' algorithm which mimics
the back and forth reptation motion of chains, without explicitly
attempting to model the time consuming local segment fluctuations. For
studies of chain diffusion in a dense multi-chain system, rapid
relaxation was observed \cite{DEUTSCH} using this method. Other
studies, however, suggest that the slithering snake motion is prone to
self-trapping \cite{SOKAL} and that chains can become locked in
`cages' \cite{LEONTIDIS}.  Furthermore, the method is inapplicable for
tethered or non-linear polymers.

Recently there has been some effort to develop MC algorithms
specifically tailored to deal with multi-polymers systems.  One such
method has been proposed \cite{PANT} that affects large conformational
changes by splitting and bridging parts of chains. This approach is
efficient, but necessarily introduces chain-length polydispersity into
the system. Other specialised multi-chain algorithms, suitable for
lattice chains at very high densities close to or at, saturation, have
also been developed \cite{MANSFIELD,PAKULA}.  However, their general
applicability is limited and so will not be considered further here.

A more attractive class of MC algorithms suitable for multi-polymer
simulations, are {\it ab initio} `growth' algorithms, that attempt to
engineer large scale configurational changes by removing a chain from
the system and placing it elsewhere, or by replacing parts of a chain.
The previously removed chain or chain portion is reinserted into the
system with the aid of a growth algorithm. The viability of the
technique as a whole depends crucially on the manner in which this
growth is implemented. Several approaches are possible. The most basic
is to employ a simple (non-reversal) random walk in which a chain of
length $N$ is grown as a sequence of $N$ randomly oriented steps. The
construction fails if the growing chain visits any point in space that
is excluded due to the presence of other particles. In practice this
approach is almost useless (particularly for dense multi-chain
systems), since a growing chain sooner or later encounters excluded
volume. As a result, the chain construction rate falls exponentially
with $N$, i.e.  $f_{con}\simeq \exp(-c_0 N)$.

In view of the high attrition rate for the simple random walk,
Rosenbluth and Rosenbluth (RR) \cite{ROSENBLUTH} introduced a biasing
technique for increasing the chain construction rate.  In the original
formulation of this method, one considers a single chain growing on a
lattice. At the $i$th step of its construction, the chain is accorded
$k$ possible distinct random growth directions (on a lattice $k$ is
usually taken as $q-1$, with $q$ the coordination number).  Unlike the
simple random walk, however, the chain at each step does not blindly
choose a random direction from the set of $k$. Instead it examines all
$k$ possibilities and chooses the actual step from the subset of $w_i$
directions that avoid excluded volume. If $w_i =0$, then the growing
chain has run into a dead-end or `trap' and the construction fails.
Of course, choosing the actual step from the subset of $w_i$
directions instead of the full complement of $k$, for $i=1,..N$,
introduces a bias. The total construction bias accumulated over $N$
steps is subsequently compensated for in the calculation of averages.

Although considerably more efficient than the simple random walk
construction mentioned above, the RR technique is still inefficient for
very long chains due to its shortsightedness. By looking ahead only one
step, it often blunders into traps. This trapping means that the
attrition is still ultimately exponential in $N$, i.e. $f_{con}\simeq
\exp(-c_1 N)$, although with a coefficient $c_1\ll c_0$. Alternatively
the attrition can be reduced by employing the `scanning method' of chain
growth \cite{MEIROVITCH}. For each of the $k$ directions of an $i$'th
step, one constructs a Cayley tree of all free pathways of $l$ further
steps. The actual next step is then made probabilistically dependent on
the number of progeny spawned by each of the $k$ directions.  However
the CPU time expended on constructing the Cayley tree increases
exponentially with $l$, which limits the scanning method considerably.
In view of this problem, a variant of the method known as Double
Scanning \cite{DSM} has been developed in which only a subset
of all possible future paths are explored. This speeds up the 
the method, but both renders the sampling approximate and reduces
the ability to circumvent traps.

The RR algorithm, originally developed for the study of single chains,
has recently been incorporated into a multi-chain MC scheme known as
Configurational Bias Monte Carlo (CBMC)
\cite{SIEPMANN,SIEPMANN92,FRENKEL92}. Within the canonical ensemble
formulation of the CBMC scheme, one considers a system containing a
fixed number of chains, each of length $N$. A chain is chosen at
random and an attempt made to regrow it elsewhere in the system, using
the RR scheme described above. If the construction succeeds, the new
chain is accepted and replaces the old chain with a probability
dependent on the ratio of the total construction bias of the new and
old chains.  The CBMC method can be used with both lattice and
continuum polymer models \cite{FRENKEL91}.

The chief advantage of the CBMC method is that large scale
configurational changes can be performed at one go. Indeed the method
has proved itself very effective at low and moderate densities for
chain-lengths up to a few dozen monomers. Unfortunately, it transpires
that the acceptance rate for chain insertions falls exponentially with
increasing chain length. This problem is symptomatic of the fact that
the RR method produces chains having a probability distribution that
differs exponentially in $N$ from the Boltzmann distribution
\cite{BATOULIS}. Taken together with the exponential attrition rate
for chain construction, it is clear that the method will be
inefficient when dealing with long chains and/or very dense systems.
In such situations, the best that can be done is to try to regrow
terminal or intermediate portions of chains
\cite{DIJKSTRA,VENDRUSCOLO}.

In this paper we propose a new MC method suitable for dense
multi-chain systems.  Our approach is close in spirit to CBMC, but
instead of using the standard RR algorithm to grow chains, we employ a
more sophisticated biased growth technique which we call `recoil
growth'. The strength of the recoil growth method resides in its
ability to avoid traps and excessively dense regions by deploying a
retractable `feeler' of maximum length $l$, to probe the territory
ahead of the growing chain. This feeler recoils from traps and dense
regions, and thus guides the chain along favourable pathways. The
result is a substantial improvement (compared to CBMC), in chain
construction and acceptance rates at high density and long
chain-lengths. Such a recoil idea has been proposed by Alexandrowicz
and is described in a recent publication \cite{ALEXANDROWICZ}.  The
present application of recoil is however different from that of
ref.\cite{ALEXANDROWICZ}. In the present version, a fully grown new
chain replaces an old one with the help of a MC accept/reject step
determined by their respective weights, just like in CBMC. In contrast
ref.\cite{ALEXANDROWICZ} utilizes a sequence of stepwise lotteries `on
the go', which are applied as the new chain is being grown and are
determined by its own weight alone. Such a procedure involves a
certain approximation, (ie. it is not exact) but is substantially more
efficient.

The layout of our paper is as follows. In section~\ref{sec:back} we
describe our recoil growth algorithm in detail and show how it may be
incorporated within a MC scheme that fulfills the detailed balance
condition. Then, in section~\ref{sec:results}, we test the validity of
the method and investigate its properties via extensive simulations of
self avoiding walks on a cubic lattice. A comparison of efficiency
with the CBMC method is also made. Finally section~\ref{sec:concs}
summarizes and discusses our findings.

\section{Description of the algorithm}

\label{sec:back}

The form of our proposed recoil growth scheme is motivated by the need
to address the two principal deficiencies of the RR growth method. The
first deficiency is the inability to avoid traps.  These traps have a
wide distribution of depths \cite{BATOULIS,HEMMER}, and owing to its
myopia, the RR algorithm is incapable of avoiding them. Since the
number of traps increases rapidly with density, the RR algorithm thus
suffers low construction rates, particularly for long chains.

The second deficiency relates to the distribution of chains produced
by the RR algorithm. In order to avoid excluded volume and thus
maintain a relatively high construction rate, it is necessary to
employ a rather large number of alternative growth directions $k$.
Unfortunately, this leads to an indiscriminate growth procedure that
yields an ensemble of low weight chains. As a result, the acceptance rate
for CBMC moves falls exponentially with chain length.  Clearly
therefore any proposed improvement to the RR growth algorithm must
strive to improve both the construction rate {\em and} the acceptance
rate.  Our recoil growth algorithm achieves this by positioning at the
head of the growing chain, a long retractable feeler having the
ability to recoil from traps and excessively dense regions. The
advantage comes from the fact that if the growth procedure encounters
a trap or a dense region, we do not terminate the growth of the whole
chain, but merely recoil back the required number of steps (up to a
maximum length $l$), and try elsewhere. In this way the growing chain
both avoids traps and finds better pathways through the system.

In what follows we describe how the recoil growth scheme can be
incorporated into a canonical MC scheme in which individual chains are
regrown within the system.

\subsection{Chain construction.}

\label{sec:constr}

A new chain is generated according to the recoil growth procedure as
follows. To begin, the first monomer (denoted by $i=1$) is placed
at random in the system. From this first monomer one attempts to grow
a single step (bond) to the second ($i=2$) monomer. For this purpose a
maximum of $k$ distinct directions are permitted. One chooses a random
direction and attempts to grow the chain a single step. If this growth
is blocked, however, then another direction is chosen, repeatedly if
necessary, up to the maximum of $k$ attempts. Possible reasons for
blockages are excluded volume interactions with previous monomers of
the chain, or with other chains. If at the $b$th attempt no excluded
volume is encountered, we place the second monomer on the lattice,
record the number of unused directions $k-b_1$ and proceed to grow the
next step. The same process is repeated for all subsequent monomers
$i=1,...N$: the chain grows at the $b_i$ attempt and the number of
unused directions $k-b_i$ are recorded for all $i$.

Suppose, however, that at the $i$th monomer, the chain failed to grow to
the length $i+1$ within the maximum of $k$ possible attempts. In this
case it returns to the $i-1$ monomer and renews its attempts to grow
from there, using the $k-b_{i-1}$ previously unused directions.
Similarly if it fails to grow a step at the $i-1$ monomer (within the
total number of $k$ attempts), it falls back to the $i-2$ monomer. In
difficult situations the chain may thus repeatedly grow and fall back.
However it is not permitted to fall back indefinitely. If $i_{max}$ is
the {\em greatest} length that the chain attained in its whole growth
history, then it may fall back no further than to the length
$i=i_{max}-(l-1)$. If the chain does recoil to this length (or if $n<l$
and it recoils to its starting point), and still fails, then the growth
process terminates.  We note that once the chain has grown up to the
maximal length $i_{max}$, monomers $i=1,...,i_{max}-(l-1)$ are fixed and
no longer subject to recoil growth. This `fixed chain' only advances by
a step when the number of steps ahead of it has attained length $l$.
Chain monomers $i> i_{max}-(l-1)$ may be thus regarded as constituting a
{\em retractable feeler of maximum length $l$}, which allows the growing
chain (of fixed monomers) to `look ahead' by $l$ steps.  In this sense
our scheme bears some resemblence to the  Double Scanning Method of
Meirovitch \cite{DSM}, although as discussed further in
section~\ref{sec:concs}, there are important differences.

An example of a possible recoil growth scenario is illustrated schematically
in figure~\ref{fig:tree}(a).  The growth procedure can be envisaged as
taking place on a underlying random tree that is revealed as the
growth proceeds.  The tree is formed by assigning random directions to
the monomers. This concept of a tree will be used below for
demonstrating that the method obeys the detailed balance condition.
Should the new chain attain the desired length $N$, then it becomes a
candidate for replacing some randomly chosen existing chain, and one
proceeds to calculate the weights of both chains.

\subsection{Weight calculation}

In order to incorporate the recoil growth scheme within a MC framework
it is necessary to obtain the probability of generating a particular
chain configuration on the underlying random tree, since this quantity
enters into the detailed balance condition described below.  For
convenience we shall work with the inverse of this probability, which
we refer to as the chain `weight'. The calculation of this weight
proceeds with the use of feelers. Since the procedure for weight
calculation differs slightly for the candidate (new) chain and the
existing (old) chains, we shall describe them separately.

\subsubsection{Candidate (new) chain}

One visits the successive monomers of the new chain and from each one
attempts to grow (in turn) feelers of length $l$, exploring the
$k-b_i$ first step directions that remained untried in the
construction process (a feeler already exists along the chain backbone
itself).  These feelers are grown using exactly the same recoil growth
procedure as employed for chain construction. Thus they are allowed to
retreat if necessary, because of blockage by monomers belonging to the
same feeler, to the candidate chain (up to where the feeler starts),
and to other chains.  However if a feeler recoils by $l$ steps back to
its starting point, it is deemed to have failed; successful feelers
attain the length $l$. (For the last $i=N-l+1 \cdots N$ monomers, the
feelers are progressively shortened by one step). Examples of feeler
growth scenarios are given in figure~\ref{fig:tree}(b).  The weight
(inverse construction bias) at each step of the new chain is given by
the number of alternative directions in which the $i$th to $i+1$th
step could have been taken under the rules of our recoil growth scheme
(we shall elaborate on this point at the end of this section). The
total chain weight is therefore

\begin{equation}
\label{eq:weight}
W_n = \prod_{i=1}^{N-1}\frac{w_i}{k} ,
\end{equation}
where $w_i$ is the number of successful feelers starting at monomer
$i$ (including the one along the existing chain) and for convenience
we have normalised by $k$.

\subsubsection{Existing (old) Chain}

When an existing chain is selected for possible replacement by the
candidate new chain, its weight $W_o$ is computed by constructing new
feelers along $k-1$ directions randomly assigned to the monomers (the
direction along the existing chain is of course not examined).
Otherwise the calculation of $W_o$ proceeds exactly as for the
candidate new chain i.e. with the help of eq.~\ref{eq:weight}.

\subsection{Detailed balance and acceptance probability.}

In a MC phase space trajectory, microstates are visited with the
appropriate equilibrium probability distribution if one imposes the
detailed balance condition, which demands equality of the transition
rates from some initial (old) state $o$ to some final (new) state $n$,
and back again, i.e.

\begin{equation}
\label{eq:db}
q_nP(n\rightarrow o) = q_oP(o\rightarrow n),
\end{equation}
where $q$ is the equilibrium probability (Boltzmann) distribution to
which the old and new states belong, and $P(n\rightarrow o)$ is the
transition probability from state $n$ to state $o$.

Our canonical ensemble Monte Carlo procedure involves attempting to
exchange an old chain with a new chain. For the purposes of
demonstrating detailed balance, one can imagine this to occur with the
help of random trees as follows.

\begin{enumerate}
  
\item[(i)] A random tree $t_n$ is generated having $k$ bifurcations at
  each step, cf. fig~\ref{fig:tree}.
  
\item[(ii)] The recoil growth process is executed upon the random tree
  $t_n$ and a new chain configuration is generated.
  
\item[(iii)] A random tree $t_o$ is generated around the old chain,
  such that the old chain configuration lies on the tree.

\end{enumerate}
As we shall see, the last step (iii) is necessary so that one can
define a reverse Monte Carlo move.

Clearly within this formulation, the probability of generating given
random trees for both the new {\em and} old chains enters the
expression for the transition probability, which is given by:

\begin{equation}
\label{transprob}
P(n\rightarrow o) =\sum\limits_{t_o,t_n}
P_g(c_n,t_n)P_g(t_o|c_o)P_a((c_o,t_o)\rightarrow (c_n,t_n))
\end{equation}
Here, $P_g(c_n,t_n)$ is the probability of generating the new chain
configuration $c_n$ on the tree $t_n$ with the recoil growth
algorithm. This may be writtem $P_g(c_n,t_n)=P_g(t_n)P_c(c_n\mid t_n)$
where $P_g(t_n)$ is the {\it a priori} probability of generating a given
tree $t_n$ and $P_c(c_n\mid t_n)=1/W_n$ is the probability of
constructing the new chain on that tree.  $P_g(t_o|c_o)$ in
eq.~\ref{transprob} is the probability of generating the old tree
$t_o$ given the old chain configuarion $c_o$.  The probability of
accepting the exchange of chains ($o\to n$) is given by
$P_a((c_o,t_o)\rightarrow (c_n,t_n))$, and the sum extends over all
possible combinations of new and old trees. The probability of the
reverse move $P(n\rightarrow o)$ is simply obtained by subsitituting
$n$ by $o$ and vice-versa in eq.~\ref{transprob}.

Detailed balance (eq.~\ref{eq:db}) is satisfied if one imposes the
stronger condition of `superdetailed balance'
\cite{FRENKEL91,DAANSBOOK}, namely that microscopic reversibility is
fulfilled for {\em every} particular choice of random trees $t_o$ and
$t_n$.  If we set the tree generation probability $P_g(t)$ to be
uniform for all trees, one readily finds that the new state should be
accepted with a probability:

\begin{equation}
\label{eq:accept}
P{(o\to n)}=\min\left (1,\frac{q_oW_n}{q_nW_o}\right).
\end{equation}
This is the acceptance probability used in the Metropolis step of the
MC scheme.

\subsection{Remarks on the algorithm}

Having completed our definition of the recoil growth MC method, the
following remarks are appropriate.

The use of the random tree concept in the detailed balance condition,
simply constitutes a generalisation of the concept of a specific
random choice of possible one-step directions, which was introduced in
ref. \cite{FRENKEL91} to demonstrate the validity of continuum CBMC.
Indeed, for feeler length $l$=1, our method simply reduces to CBMC.
However, the tree concept also serves to clarify a somewhat subtle
point concerning the way we have formulated the construction algorithm
in section~\ref{sec:constr}. Clearly an alternative chain construction
procedure would be to attempt to grow at each step $j$ of the fixed
chain, {\em all} $k$ feelers, and to choose the $j+1$ monomer randomly
from among the set of first step directions of the successful feelers.
However, the picture of chain growth on an underlying random tree
shows that since all feeler directions are chosen randomly, it is
permissible (and indeed much more computationally efficient) to work
with just one feeler at a time during construction. Thus the fixed
chain grows from the $j$ to $j+1$ monomer along the first step of a
successful feeler.  The remaining steps of this feeler then constitute
an incomplete feeler of length $l-1$ with respect to the $j+1$
monomer.  One then proceeds to attempt to extend this feeler to length
$l$ using the normal recoil growth prescription.  If successful, the
chain grows to the $j+2$ step along the first step of the feeler and
the process repeats. Only if the existing feeler cannot be extended to
length $l$ need one attempt to grow another feeler from scratch.  Of
course, if the chain successfully attains length $N$, it is necessary
to return and attempt to grow the remaining feelers $k-b_j$ from each
monomer in order to calculate the weight of the candidate new chain.

With regard to the chain weight $W$ defined in eq.~\ref{eq:weight}, we
note that this is simply the product of the number of allowable
directions $w_i$ in which the chain could have grown at each step.
However, it is important to appreciate that the {\em criterion} for
what constitutes an allowable direction is arbitrary--it effects only
the efficiency and is irrelevant to detailed balance. In our recoil
growth scheme, the criterion adopted requires that an allowable first
step direction may be continued for $l-1$ further steps, such that the
entire sequence avoids excluded volume. Therefore the number of
surviving feelers $w_i$ at each monomer constitutes the number of
single step directions (out of the maximum number of $k$) that satisfy
our criterion. Accordingly, the weight $W$ defined in
equation~\ref{eq:weight} is the total inverse bias of a chain's
construction and by compensating for $W$ in the acceptance
probabilities, we ensure that detailed balance is obeyed. Of course,
other criteria for the number of allowable directions at each step are
also possible. Less stringent than ours is the RR criterion of CBMC,
which requires that an allowable growth direction be able to continue
for only one step (i.e. $l=1$).  Still more lenient is the
construction of self avoiding chains with the help of the simple
random walk which allows all directions, including those that do
encounter excluded volume. All three methods satisfy detailed balance,
but they calculate $W$'s that increase respectively, and pay the price
in decreasing construction efficiency. Thus our intricate recoil
growth procedure is merely an information gathering device that
enables us to make a `good' decision in the sense of efficient
sampling.
 
Finally, we note that if we make $k$ smaller than the coordination
$q-1$, our calculation of weight $W$ for a given chain configuration
is stochastic in the sense that a feeler's survival depends to some
extent on the random choice of the $k$ directions at each step of the
$l$-step feelers. Thus sometimes the weight of a chain will be
underestimated, while at other times it will be over-estimated with
respect to a measurement using the full complement $k=q-1$.  On
average however, one expects that the correct distribution of $W$'s
will be produced. Stochastic sampling of $W$ is also unavoidable in
the continuum implementations of the CBMC method \cite{FRENKEL91}.
Incidentally, $k$ need not be an integer.  Thus, for example, $\langle
k \rangle=3.5$ is obtained by half the time assigning $k=3$ and half
the time assigning $k=4$. For long chains we show below that the
algorithm's efficiency can be rather sensitive to the choice of
$\langle k \rangle$.

\section{Results}
\label{sec:results}

To study the properties of the proposed recoil growth MC method we
have performed canonical ensemble simulations of self avoiding walks
on a simple cubic lattice. We begin by describing tests to establish
the validity of the method and then proceed to examine its
characteristics with respect to chain construction and acceptance.
Finally we assess the methods efficiency at relaxing the sample and
compare it to that of the CBMC algorithm.

Our first task was to verify that the method produces chains with the
correct statistical properties. This was achieved by means of a
comparison with the CBMC method for the following two sets of chain
length and density consisting of 100 chains:

\begin{enumerate}
  
\item[(i)] $N=100$ at a monomer density of $\rho=0.1$
  
\item[(ii)] $N=40$ at a monomer density $\rho=0.6$

\end{enumerate} 

These are conditions for which CBMC operates reasonably well and
should therefore supply bountiful statistics.  A large number of
independent equilibrated system configurations were generated using
both the CBMC method $(k=5, l=1)$, and the recoil growth method with
$k=3, l=5$.  For both schemes we monitored the distributions of two
quantities, namely the square of the gyration radius $R_g^2$, and the
logarithm of the Rosenbluth weight $W_{RR}$ of the chains. This latter
quantity is defined as $W_{RR}=\prod_{i=1}^Nw_i/k$, where $w_i$ is the
number of free first-step continuations from the $i$th monomer,
including the step of the main chain itself.  The results for system
(i) are shown in figure~\ref{fig:RRWHT}. Clearly there is a high
degree of accord between the CBMC and recoil growth methods with
respect to the distributions of $R^2_g$ and $\ln W_{RR}$. For the
averages, we obtain $\langle R^2_g\rangle=32.25(4)$, $\langle\ln
W_{RR}\rangle=-8.252(8)$ for CBMC and $\langle R^2_g\rangle=32.22(4)$,
$\langle\ln W_{RR}\rangle=-8.263(12)$ for recoil growth.  For system
(ii) we obtain for the averages of the distributions $\langle
R^2_g\rangle=11.120(26), \langle\ln W_{RR}\rangle=-19.86(6)$ for CBMC
and $\langle R^2_g\rangle=11.136(13), \langle\ln
W_{RR}\rangle=-19.84(3)$ for recoil growth.

We now turn to an examination of the $N$ dependence of the chain
construction and exchange acceptance rates $f_{con}(N)$ and
$f_{acc}(N)$ respectively for various values of $\langle k\rangle$ and
$l$. Later we will see that for given choice of $N$ and $\rho$, a
relatively fast search locates compromise values of $\langle k\rangle$
and $l$ which ensure the good performance of the recoil growth method.
At present, however, we wish to elucidate systematically the role
played by these parameters. To this end it is instructive to fix the
feeler length at a moderate value of $l=5$ and consider the effect of
varying $k$.

The results for $f_{con}(N)$ are presented in fig.~\ref{fig:rates}(a).
They show that if the construction utilizes the full complement of
directions $k=q-1$, the growing chain avoids almost all traps and we
find $f_{con}(N)\approx 1$ over the entire range of $N$ studied.  As
$\langle k\rangle$ decreases it still suffices to compensate the
average loss of continuations due to excluded volume, but owing to
fluctuations $f_{con}(N)={\rm const}\alt 1$. Figure~\ref{fig:rates}(a)
shows that this is indeed the case for $k=5,3$. However, if we set
`starvation' values of $\langle k\rangle =2.0$, or $\langle k\rangle
=1.8$, which do not compensate for the average loss due to excluded
volume, we pass to a strong exponential attrition.  The sharpness of
the transition between the compensation and the starvation regime
makes it easy to determine the marginal value of $k$ that ensures that
excluded volume is just compensated.

Having demonstrated that for a given moderate $l$, a large $\langle k
\rangle$ helps $f_{con}$, we now show that a minimal $\langle k
\rangle$ close to (but not within) the starvation regime serves to
benefit the exchange acceptance rate, $f_{acc}$. As mentioned in the
preceeding section, a large $\langle k \rangle$ leads to an
indiscriminate growth that yields many low quality chains having
relatively small weights. This implies a random sampling
which differs exponentially from the correct Boltzmann distribution,
which in turn leads to an acceptance rate $f_{acc}(N)$ that falls
exponentially with $N$. However a minimal value of $\langle k \rangle$
causes the construction to recoil from relatively dense regions, and
grow anew along a less difficult path, thus producing `higher quality'
(relatively large weight) chains.  Figure~\ref{fig:rates}(b)
shows $f_{acc}(N)$ for $l=5$ at various $\langle k \rangle$.  One sees
that although the acceptance rate still decreases approximately
exponentially with chain length, it improves markedly as $k$
decreases.

It transpires, however, that in many cases $f_{acc}$ is a highly
misleading indicator of the sampling efficiency because the majority
of chain exchanges are confined to small distinct regions of the
system, leaving the other chains mostly untouched. The problem is
traceable to the fact that some chains (those having low weight) are
easily removed from the system. When such a chain is removed, a large
vacancy is created into which another chain will preferentially grow
shortly afterwards.  Since the new chain is necessarily highly
correlated with that it replaced, it too is quickly removed. This
creates the illusion of a brisk turnover of chains, while in fact the
configuration as a whole remains substantially unaltered.  This effect
has also been recently noted by other authors in connection with the
CBMC method \cite{LEONTIDIS,SHELLEY}.

In view of the unreliability of $f_{acc}$ for indicating the true rate
of relaxation, we have adopted an alternative measure of efficiency,
namely the effective chain turnover rate, defined as the CPU time
required to replace $95\%$ of some arbitrary starting sample of
chains. Optimal values of $\langle k\rangle $ and $l$ are thus those
that maximize this effective turnover rate, or indeed any other
suitably chosen time autocorrelation function. In fact, in practice
this optimization is not difficult to achieve and suitable values can
be readily gauged from very short runs on small systems. In most cases
we find that a choice of feeler length in the range $3\alt l \alt 8$
is a good starting point.  The optimal choice of $\langle k\rangle$
depends on the density (and to a much smaller degree on the choice of
$l$). Larger $\langle k\rangle$ values, approaching the coordination
number are required at higher densities. The sensitivity of the
efficiency to the assignment of $\langle k\rangle$ increases with
chain length, so that for short chains ($N\approx 40$) $\langle
k\rangle$ should be correctly estimated to within $\pm 0.5$, while for
longer chains ($N\agt 100$) the value should be chosen to within $\pm
0.1$.  It should be pointed out that since the optimal $\langle
k\rangle$ value is often much less than the lattice coordination
number $5$, the cost of growing feelers is surprisingly low. This
point is illustrated in figure~\ref{fig:ftime} where we plot for a
number of $\langle k\rangle$ values the CPU time
$t_{feel}(l)$ (normalised with respect to CBMC) to grow feelers of
length $l$ (for $N=30$ $\rho=0.5$). One sees from the figure that for
small $l$ the time rises at most linearly in $l$ (for large $l$ the
curves round off due to the progressive shortening of the feelers as
the chain end is approached); in general we expect the time expended
on feeler growth to increase approximately linearly in their length,
with a slope that decreases markedly with $\langle k \rangle$.

To demonstrate that recoil growth method leads to net efficiency gains
we have studied the decay of the autocorrelation function of the
radius of gyration as a function of CPU time for a system of 100
chains of length $N=40$ at density $\rho=0.6$. Figure
\ref{fig:corr_rg} shows that under these conditions, the efficiency of
the recoil scheme is some three times higher than CBMC.  As another
example we have studied the rate at which the method replaces a given
equilibrated starting configuration of $100$ chains each of length
$N=100$ at density $\rho=0.5$. Figure~\ref{fig:pull} shows the
fraction of chains replaced as a function of CPU time for various
values of $\langle k \rangle$ and $l$ as detailed in the caption.  We
find that the effective chain turnover rate is maximised for a choice
$\langle k \rangle=2.7$, $l=10$, being some 6 times faster than CBMC.
A more comprehensive comparison of the relative efficiency of recoil
growth and CBMC is presented in figure~\ref{fig:eff}, which shows the
relative efficiency of the two methods in achieving $95\%$ turnover as
a function of chain length $N$ for three different densities
$\rho=0.3, 0.5$ and $0.7$. One sees that at high densities and large
chain lengths, optimised recoil growth is more efficient by up to a
factor of $50$. In Table(\ref{tab:param}) the parameters of the recoil growth
used to construct figure~\ref{fig:eff} are presented.

\section{Discussion and conclusions}

\label{sec:concs}

In summary we have proposed a new simulation method for dense,
long-chain multi-polymer systems based on `recoil growth'.  If a chain
fails during the construction due to excluded volume, it is permitted
to recoil a step and try to grow anew in another direction up to a
total of $k$ trials at each step.  If still unsuccessful it may recoil
a further step etc, up to a maximal predetermined length $l-1$.  The
construction bias is compensated for in the chain insertion/deletion
stage in the same manner as for CBMC.  The recoil allows the chain to
avoid traps of length up to $l-1$ and hence benefits the construction
rate $f_{con}$. However, it also allows the chain to avoid excessively
dense regions of the system, growth into which would otherwise produce
low quality (low weight) chains that fail the accept/reject
lottery.  Thus recoil also benefits $f_{acc}$, an effect that becomes
more pronounced as the number $k$ of directions is reduced to the
barest minimum that still yields a reasonable construction rate
$f_{con}$.  The time expended on feelers also decreases markedly with
decreasing $k$, making it possible to find (by means of a rapid
preliminary search), values of $k$ and $l$ that optimize the
relaxation rate and thus yield a substantial improvement in CPU time
efficiency over CBMC.  In this way the method extends the range of
densities and chain lengths that can effectively be studied.

Although the major factor in the faster relaxation of recoil growth is
its ability to build chains that more closely match the Boltzmann
distribution, one may speculate that the stochastic nature of the
weight determination procedure also plays a role in this regard.
Random downward fluctuations in the weight assignments to old chains
may help to dislodge high-weight chains which would not otherwise be
exchanged using CBMC.  Additionally, the non-local sampling resulting
from use of long feelers may mean that relaxation (chain exchanges) in
one area of the system influences the weight calculations for other
quite distant chains, thus promoting their exchange too.

It is instructive to compare and contrast our method with the double
scanning method (DSM) of Meirovitch \cite{DSM} since, in a sense, both
strive to achieve the same end, namely a stochastic look ahead scheme
designed to seek out favourable pathways for the growing chain. However,
the manner in which these paths are chosen differs greatly between the
two methods. In the DSM, at each step of the growing chain, a typically
large number $(50-200)$ of feelers are grown according to the RR
prescription. The next step direction of the growing chain is decided
probabilistically dependent on the proportion of feelers surviving along
each possible next step direction. In the RG scheme, however, only a
single feeler is required during the construction process and the weight
calculation differs from that of DSM in that next step directions
receive either a weight of zero or unity. The need for many RR-type
feelers to scan the future pathways is obviated by the ability of the RG
feeler to recoil from excluded volume and the fact that a successful feeler is
`recycled' at the next step of the main chain as an incomplete feeler (see
section~\ref{sec:constr}).  We therefore believe RG to be considerably
more efficient than DSM at locating favourable pathways for the growing
chain. To date, the DSM growth method has not been incoporated within an
exact multipolymer Monte Carlo scheme. 

It is interesting to note also a resemblance between our recoil growth and
an optimized enrichment algorithm \cite{WALL}.  The latter constructs
the ensemble of configurations for an isolated chain by allowing
$\langle k\rangle$ alternative growth directions at each step.  As
with recoil growth, $\langle k\rangle$ is chosen so that the number of
alternative growth directions equals the average loss per step due to
excluded volume.  Both methods require a tuning of $\langle k\rangle$
to avoid wild fluctuations of ``$W$''.  However one enrichment
construction produces a tree of $W$ alternative (correlated) chain
configurations, while recoil growth converts $W$ into the weight of a
single configuration. In fact, very recently a new MC growth scheme
based on enrichment has been proposed \cite{GRASSBERGER}.  The weights
of the chains are confined within a desired range by eliminating
probabilistically configurations with very small weight, while at the
same time enriching the sample with copies of high weight chains in
such a way that the resulting sample is unbiased.  The method allows
one to study very large chain lengths at low to moderate densities.
Unfortunately we know of no reported tests of the method at high
densities, and can therefore not compare with the recoil growth method
in this regime.

A number of extensions to the recoil growth method can also be
envisaged.  As with CBMC there should be no problem in applying the
method to continuum systems.  The effects of temperatures can also be
incorporated by introducing a threshold for feeler survival based on
its total Boltzmann weight.  The construction rate for very long
chains in the dense regime could also be improved by adopting a
`divide and conquer' approach in which one tries to regrow only short
chain sections at a time.  Although we have presented our method
within a canonical framework, it is also easy to generalise to other
ensembles such as the grand canonical (constant-$\mu VT$) or Gibbs
ensemble as has been done for CBMC \cite{SMIT,MOOIJ1}.  In such cases,
one is obliged to insert entire chains into the system (rather than
regrow chain portion as is often done in canonical simulations of long
chain systems), and the advantages of recoil growth over CBMC are
expected to exceed those of the canonical ensemble case.

Finally, there are a number of interesting physical problems to which
the recoil growth method can be applied, such as chains tethered to a
surface, or branched and star polymers.  Use of an open ensemble would
also permit the study of phase behaviour of polymer solutions and
melts \cite{WILDING}.

\subsection*{Acknowledgements}

N.B.W. and Z.A. thank Kurt Binder, Kurt Kremer, Marcus M\"{u}ller and
Peter Sollich for useful discussions. NBW acknowledges the financial
support of the EPSRC (grant GR/L91412), the Royal Society of Edinburgh
and the British Council ARC programme.  Z.A. acknowledges a Humboldt
Award. S.C. acknowledges the FOM-Institute and E.U. for financial
support [Contract No. 95032 CGP.4, Contract No. ERBFMBICT-972312]. The
work of the FOM Institute is part of the research program of `Stichting
Fundamenteel Onderzoek der Materie' (FOM) and is supported by NWO
(`Nederlandse Organisatie voor Wetenschappelijk Onderzoek'). S.C. and
D.F. thank Jonathan Doye for critical reading of the manuscript.

\subsection{Appendix}

In this appendix, we present a more formal description of the various
factors affecting the efficiency of the recoil growth method.
A similar method for CBMC has been presented in Ref.\cite{MOOIJ96}.

The success of the recoil-growth method relies on the ability to tune
the parameters $k$ and $l$ in order to obtain a good chain turnover
rate for low computational cost.  The efficiency of any MC scheme can
be estimated by ${(t_a t_{MC})}^{-1}$ where $t_a$ is the
autocorrelation time in number of MC steps and $t_{MC}$ is the CPU
time for an MC step \cite{SOKAL}. Thus the parameters of the recoil
growth scheme should be optimized to minimize the product $t_a
t_{MC}$.  The additional cost in extending the fixed chain by one
monomer is given by:

\begin{equation}\label{cost}
  \mbox{Cost}(l+1) = \mbox{Cost}(l) + C_p \times P(l) + k_{l+1}\times
  P(l) \times \mbox{Cost}(l_p)
\end{equation}
where $\mbox{Cost}(l)$ is the average cost of inserting a chain of
length $l$ and is determined by the number of calls to the subroutine
determining the energy, $P(l)$ is the probability that a chain of length
$l$ is grown, and $C_p$ is the cost of extending the existing feeler by
one step.  The third term in eq.(\ref{cost}) refers to the additional cost in
computing the weights, where $k_{l+1}$ and $l_p$ denote the number
of directions and the probe length assigned to monomer $l+1$.  The
efficiency of an insertion is then expressed as

\begin{equation}
\label{efficiency}
\mbox{Eff}(l) = \frac{P(l)}{\mbox{Cost}(l)}.
\end{equation}
Also $P(l)$ is given by $P(l+1) = P(l) \times \langle P_{add}(k_{l+1},
l_p) \rangle$ where $P_{add}$ is the probability of adding a monomer.
Using the recursive relations for $P(l)$ and $\mbox{Cost}(l)$ we can
arrive at:

\begin{equation}\label{recursive}
  \frac{\mbox{Eff}(l+1)}{\mbox{Eff}(l)} = \frac{\langle
    P_{add}(k_{l+1}, l_p) \rangle }{1+(C_p+k_{l+1}\times
    \mbox{Cost}(l_p)) \times \mbox{Eff}(l)}.
\end{equation}
In practice a number of equilibriated configurations for a polymer
system are generated. The first monomer is inserted randomly in the
system and the next is added using the recoil growth.  The
quantities $P_{add}$, $C_p$ and $\mbox{Cost}(l_p)$ are computed for
the addition of a third monomer to the chain backbone for different
$l$ and $k$.  This procedure gives an initial estimate of the parameters for
optimal efficiency with respect to insertion of chains.  It does not
include dynamical information regarding the efficient sampling of the
phase space.  The most efficient sampling is achieved when the overlap
of distributions of weights for the new and old chains is maximised.
This is achieved by choosing as small a value of $k$ as possible
consistent with a reasonable construction rate.

\clearpage

\begin{figure}[htbp]
\caption{ {\bf (a)} An example illustration of the recoil growth
  procedure described in the text, for $k=2,l=3$. The underlying
  random tree is shown as a thin dashed line.  The initial monomer is
  placed at point $O$ and the following scenario is played out: The
  growing chain first tries the path OABC, but finds it blocked. It
  recoils to point $B$ and then succeeds to attain length $l$ along
  the path $OABD$. The monomer at $A$ then becomes fixed. Construction
  continues by examining the paths $DE,DF$ which are both blocked, so
  the chain recoils again to point $A$. The open path OAG... is
  subsequently found and the second monomer is fixed at point $G$.
  {\bf (b)} For the weight calculation, the backbone of the candidate
  chain is retraced and from each monomer $i$, $b_i$ feelers are grown
  starting from the directions that remained {\em unused} during
  construction. From $O$ one feeler is grown. Path $OHIJ$ is blocked,
  and the feeler recoils to point $I$, from which it subsequently
  attains length $l$. No feeler is grown from point $A$. The
  respective weights assigned to points $O$ and $A$ are thus: $w_1 =
  2$, $w_2 = 1$. }
\label{fig:tree}
\end{figure}

\begin{figure}[h]
\caption{Histograms showing {\bf (a)} the logarithm of the Rosenbluth
  weights, for CBMC and recoil growth $(k=3, l=5$) {\bf (b)} The
  radius of gyration squared. The length of the MC run is $2\times
  10^8$ cycles.}
\label{fig:RRWHT}
\end{figure}

\begin{figure}[h]
\caption{{\bf (a)} Construction rate $f_{con}(N)$ for various values
  of $\langle k\rangle$. All simulations were carried out for $100$
  chains at density $\rho=0.5$, using $l=5$. {\bf (b)} The
  corresponding exchange acceptance rate $f_{acc}(N)$}
\label{fig:rates}
\end{figure}

\begin{figure}[h]
\caption{The feeler construction time $t_{feel}(l)$ normalised with respect to
  CBMC ($k=5,l=1$) for various values of $\langle k \rangle$, for
  $N=30$, $\rho=0.5$.}
\label{fig:ftime}
\end{figure}

\begin{figure}[htbp]
\caption{Autocorrelation function of the square of the gyration radius vs. CPU
  time for a system of $100$ chains of length $N=40$ at density
  $\rho=0.5$. The dashed line and the solid line represent the CBMC
  and the recoil growth scheme for $k=3,l=5$ respectively. The subscript $n$ in the
  average denotes that the correlation function is normalized to 1.}
\label{fig:corr_rg}
\end{figure}

\begin{figure}[htbp]
\caption{The remaining fraction of the original sample of $100$ chains
  of length $N=100$, density $\rho=0.5$ as a function of CPU time
  (secs/ALPHA).  The solid, long dashed and dotted lines correspond to
  the recoil growth with $(k,l)= (2.7, 10), (3, 10), (5, 4)$
  respectively. The short dashed line coressponds to the CBMC results.
  The inset shows the associated construction and the percentage 
  of acceptance rates on a log scale.
  The diamond, square and circle refer to recoil growth with 
  $(k,l)= (2.7, 10), (3, 10), (5, 4)$ and the triangle to CBMC.
}
\label{fig:pull}
\end{figure}

\begin{figure}[htbp]
\caption{The efficiency of recoil growth relative to CBMC, measured as
  the ratio of the CPU times required to remove $95\%$ of a given
  equilibrated starting sample of $100$ chains. Data are shown for a
  number of chain lengths at densities $\rho=0.3, 0.5$ and $0.7$. In
  each instance, the recoil growth parameters $\langle k\rangle$ and
  $l$ were chosen to yield the fastest relaxation rate.}
\label{fig:eff}
\end{figure}

%

\begin{table}[htbp]
    \caption{The $(\langle k \rangle,l)$
      values that correspond to \protect fig. (\ref{fig:eff}).}
    \label{tab:param}
\end{table}

\clearpage
\pagestyle{empty}

\begin{figure}[htbp]
  \begin{center}
    \epsfig{file=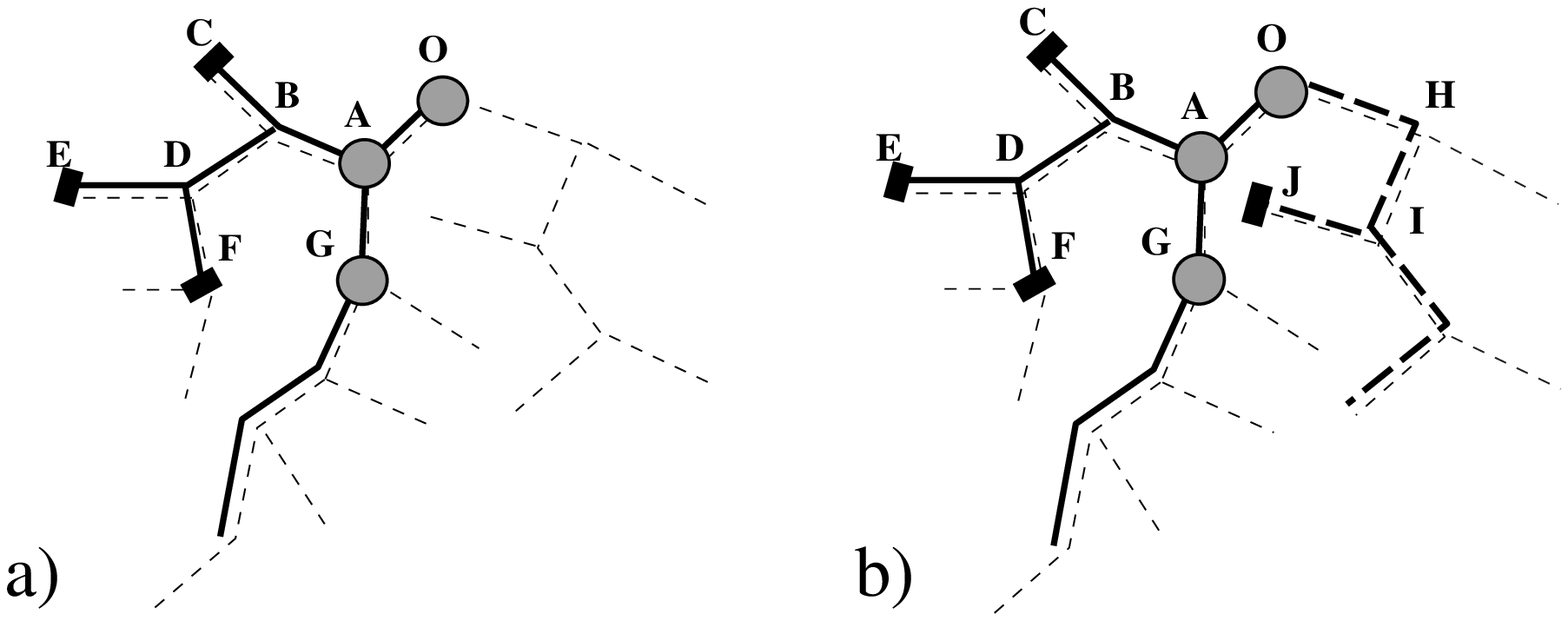,width=.9\linewidth}
  \end{center}
\end{figure}

\clearpage

\begin{figure}[h]
  \begin{center}
    \parbox[b]{0pt}{\LARGE a) \\ \vspace{1cm}\mbox{}}
    \epsfig{file=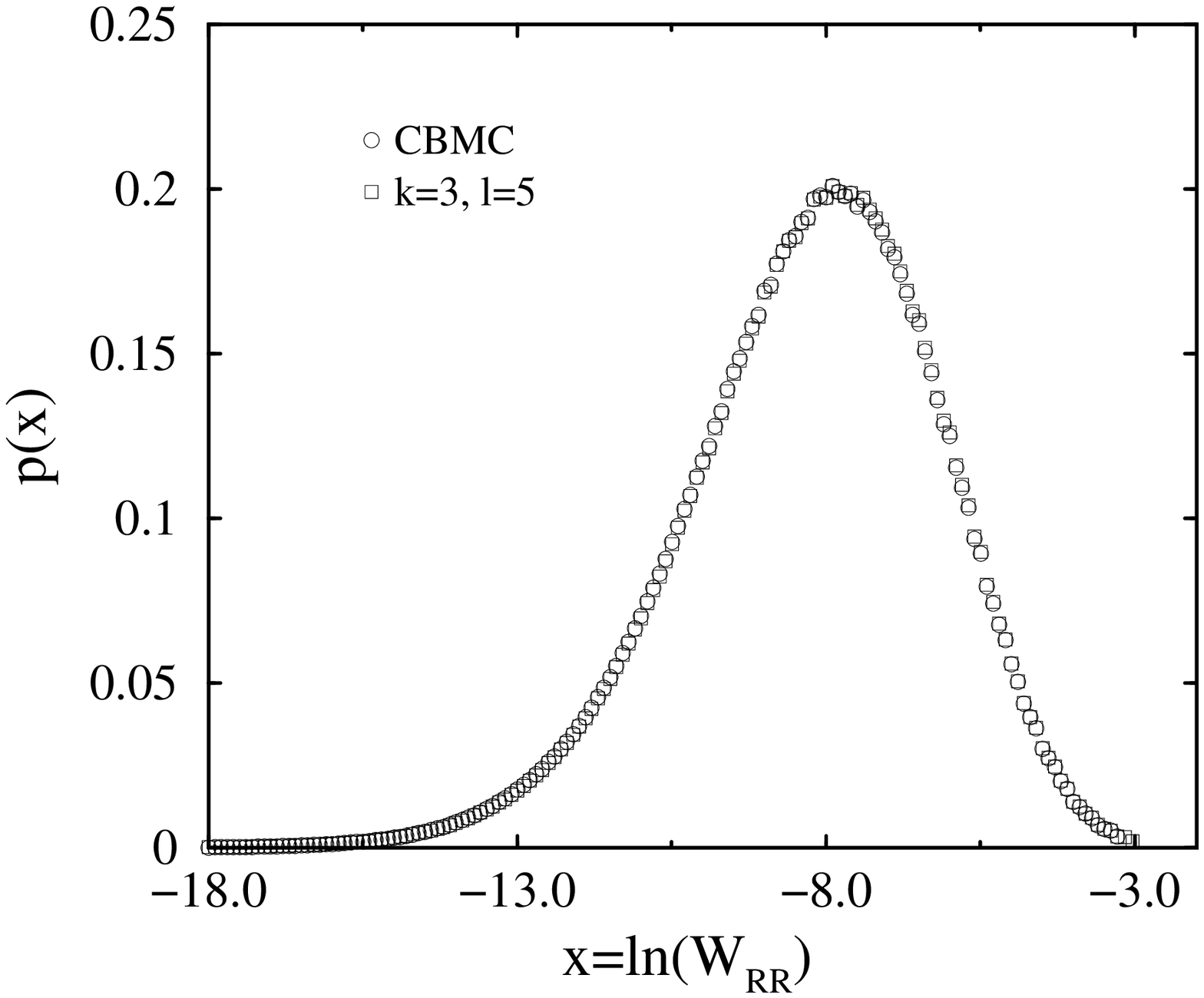,width=.8\linewidth}\\
    \parbox[b]{0pt}{\LARGE b) \\ \vspace{1cm}\mbox{}}
    \epsfig{file=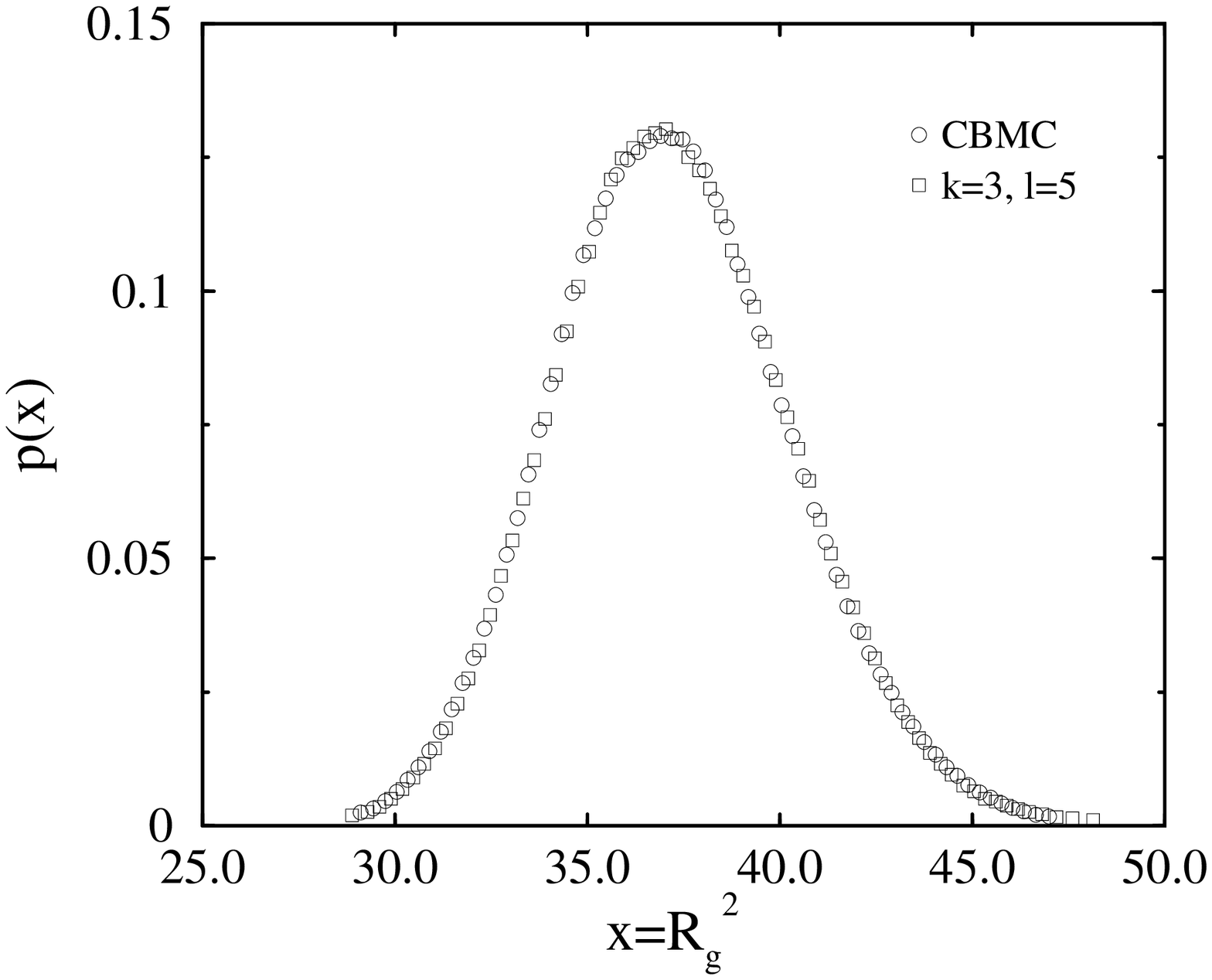,width=.8\linewidth}
  \end{center}
\end{figure}

\clearpage
\begin{figure}[h]
  \begin{center}
    \parbox[b]{0pt}{\LARGE a) \\ \vspace{1cm}\mbox{}}
    \epsfig{file=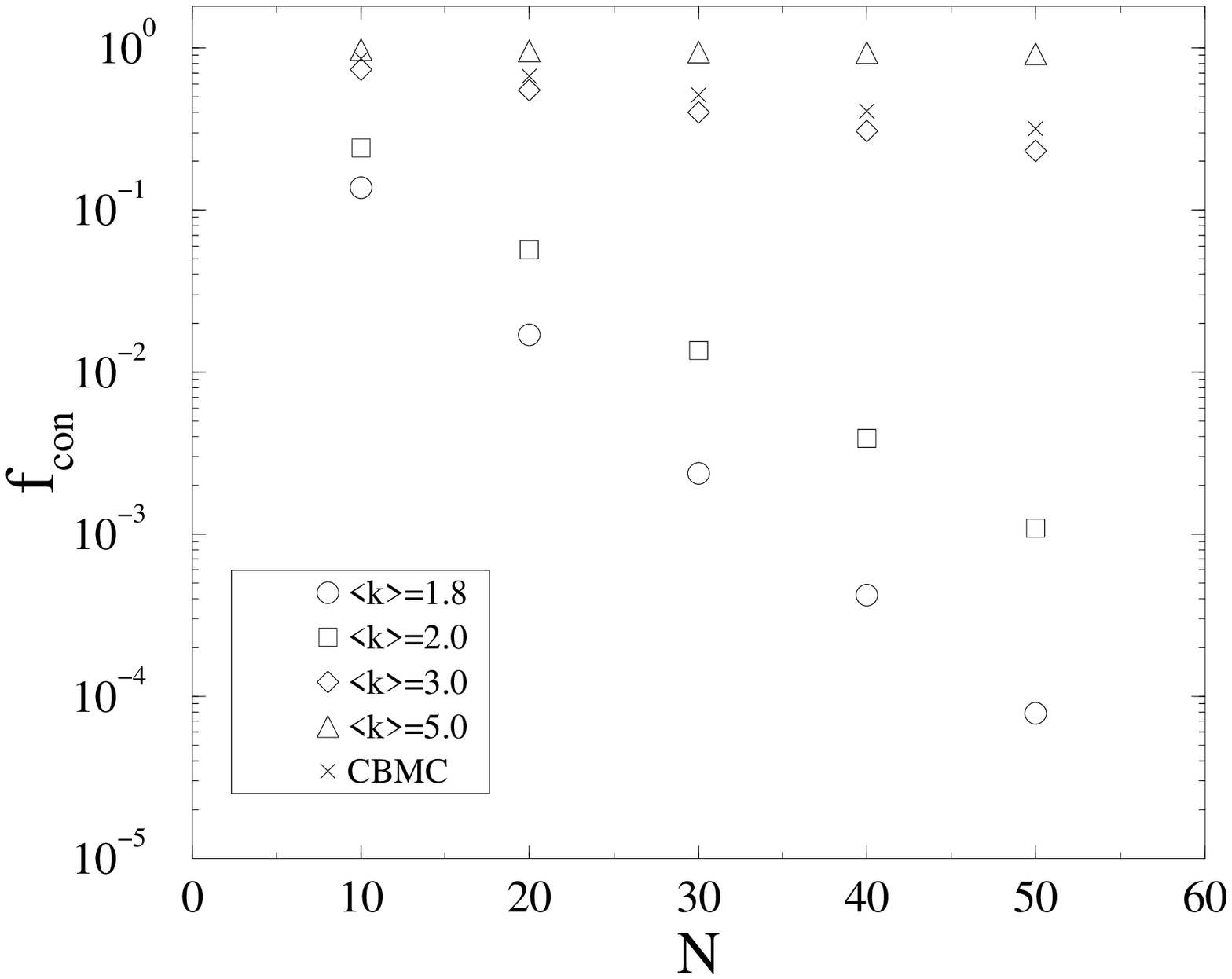,width=.8\linewidth}\\
    \parbox[b]{0pt}{\LARGE b) \\ \vspace{1cm}\mbox{}}
    \epsfig{file=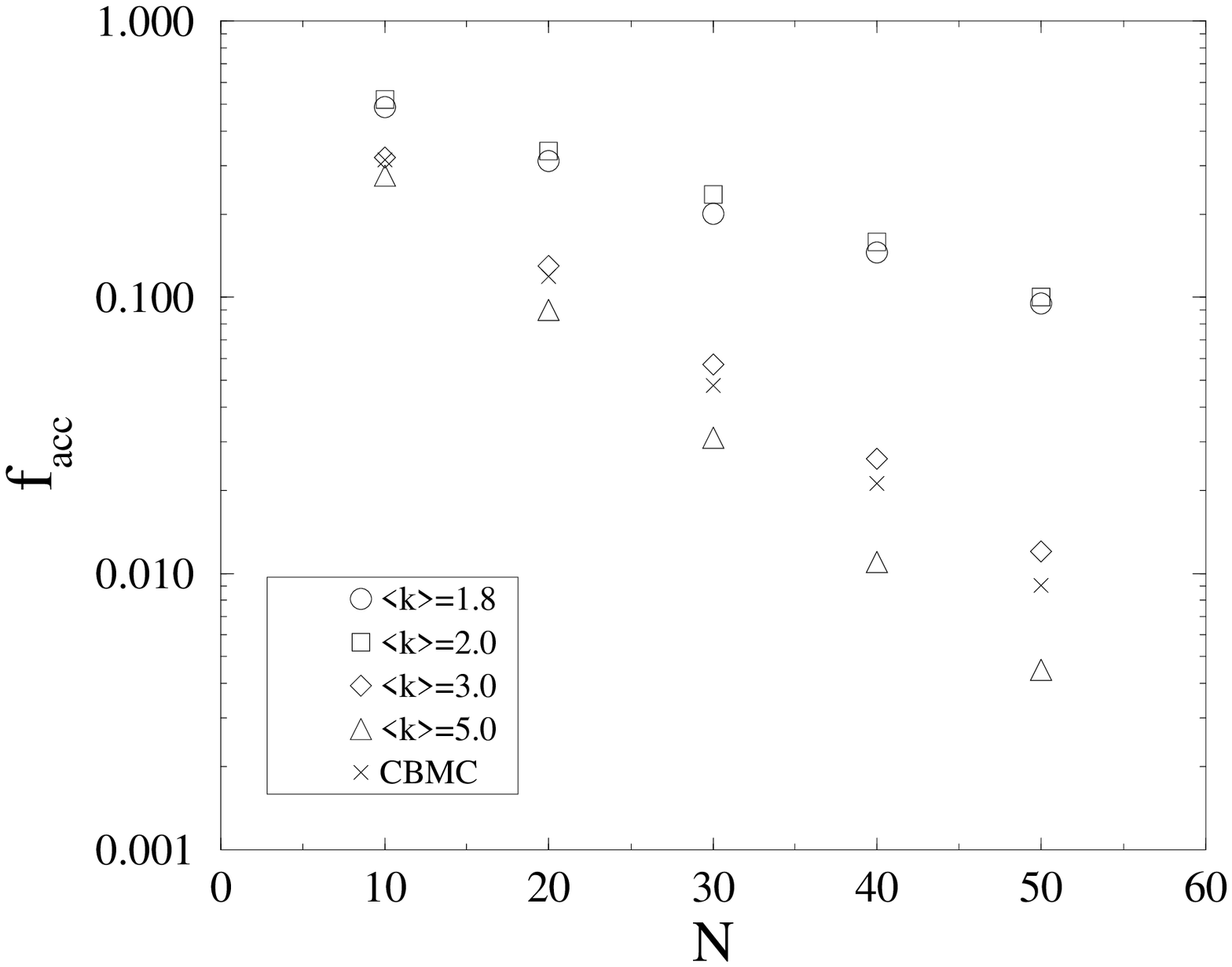,width=.8\linewidth}
  \end{center}
\end{figure}

\begin{figure}[h]
  \begin{center}
    \epsfig{file=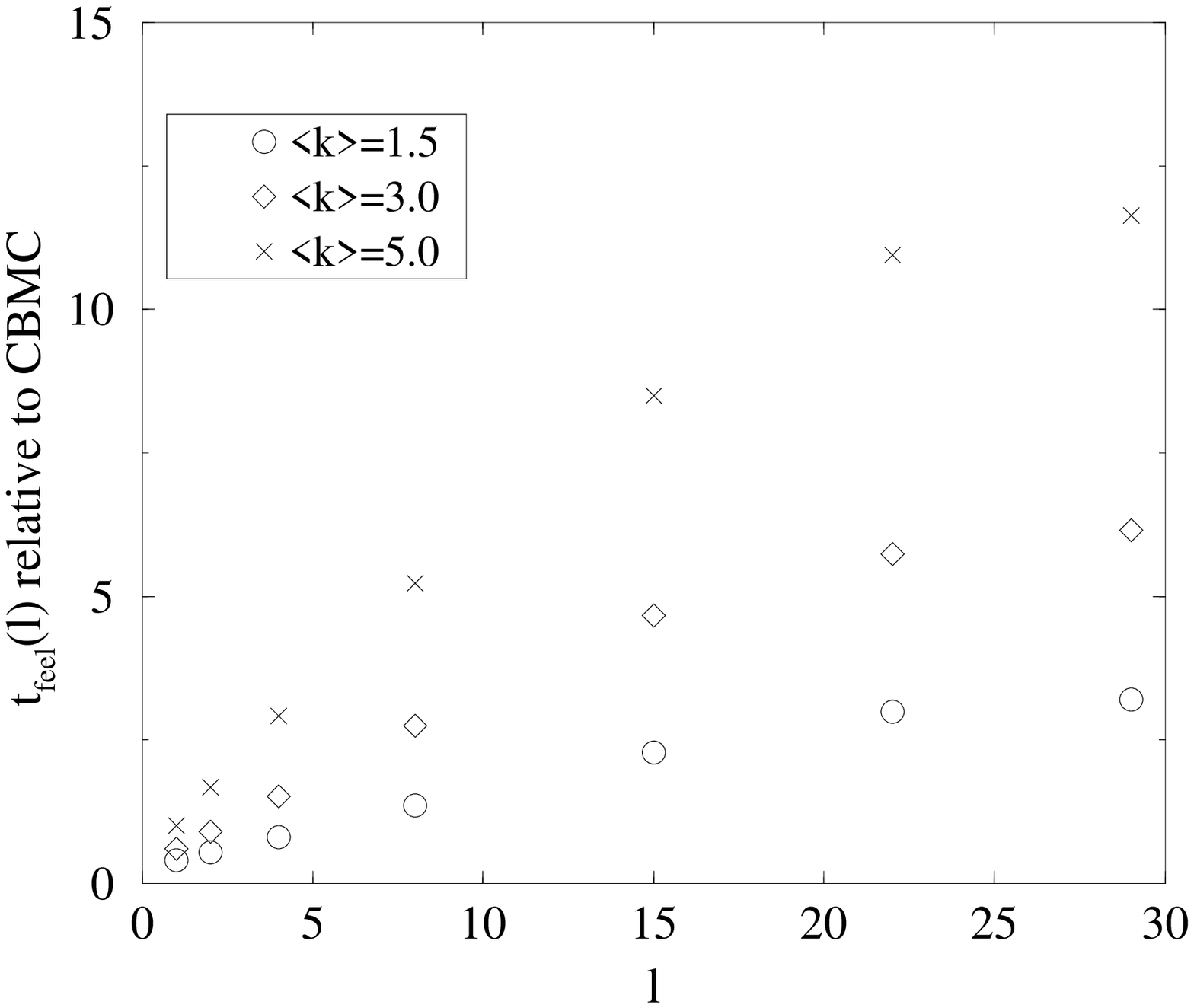,width=\linewidth}
  \end{center}
\end{figure}

\clearpage
\begin{figure}[htbp]
  \begin{center}
    \epsfig{file=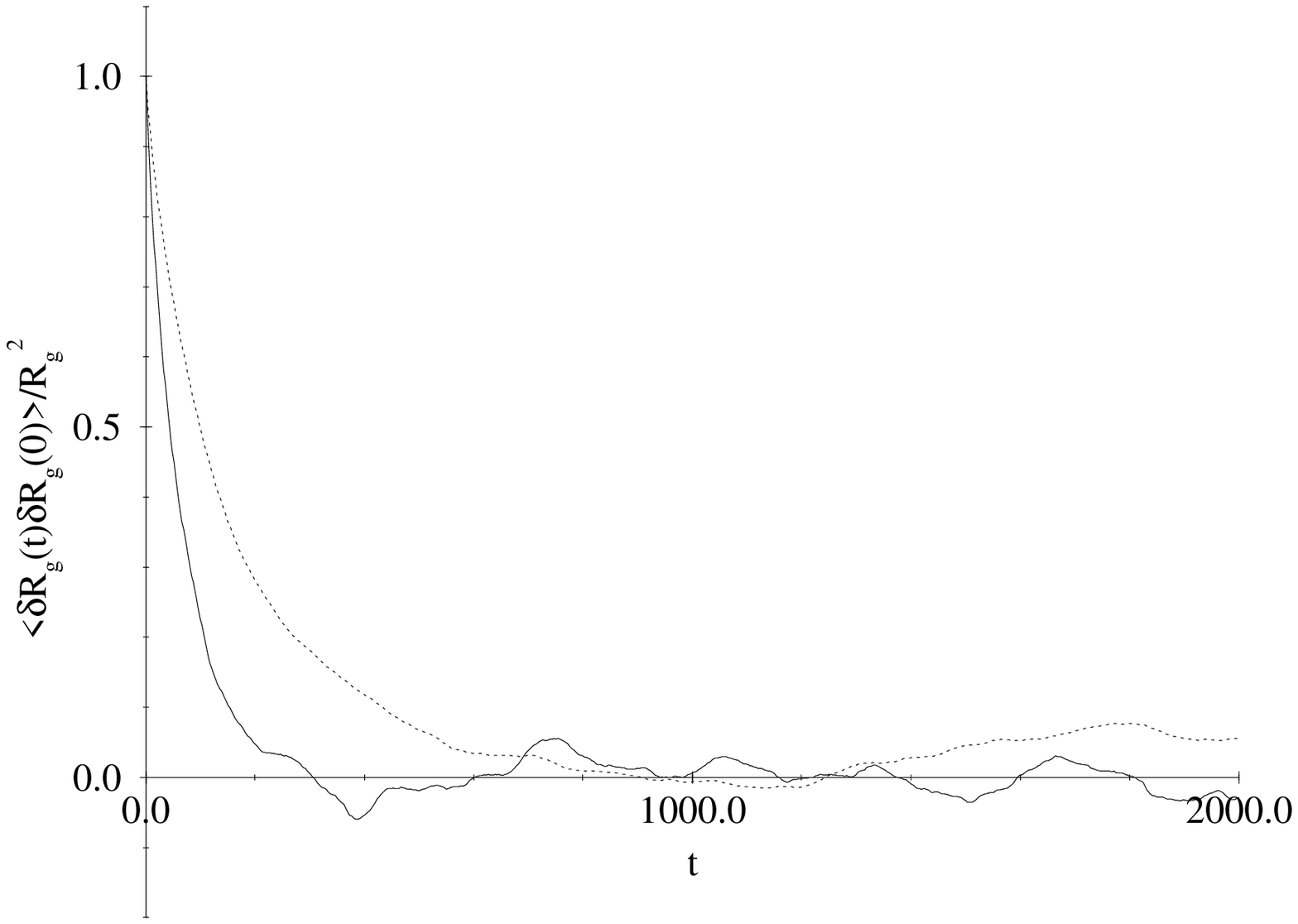,width=.9\linewidth}
  \end{center}
\end{figure}

\clearpage
\begin{figure}[htbp]
\begin{center}
  \epsfig{file=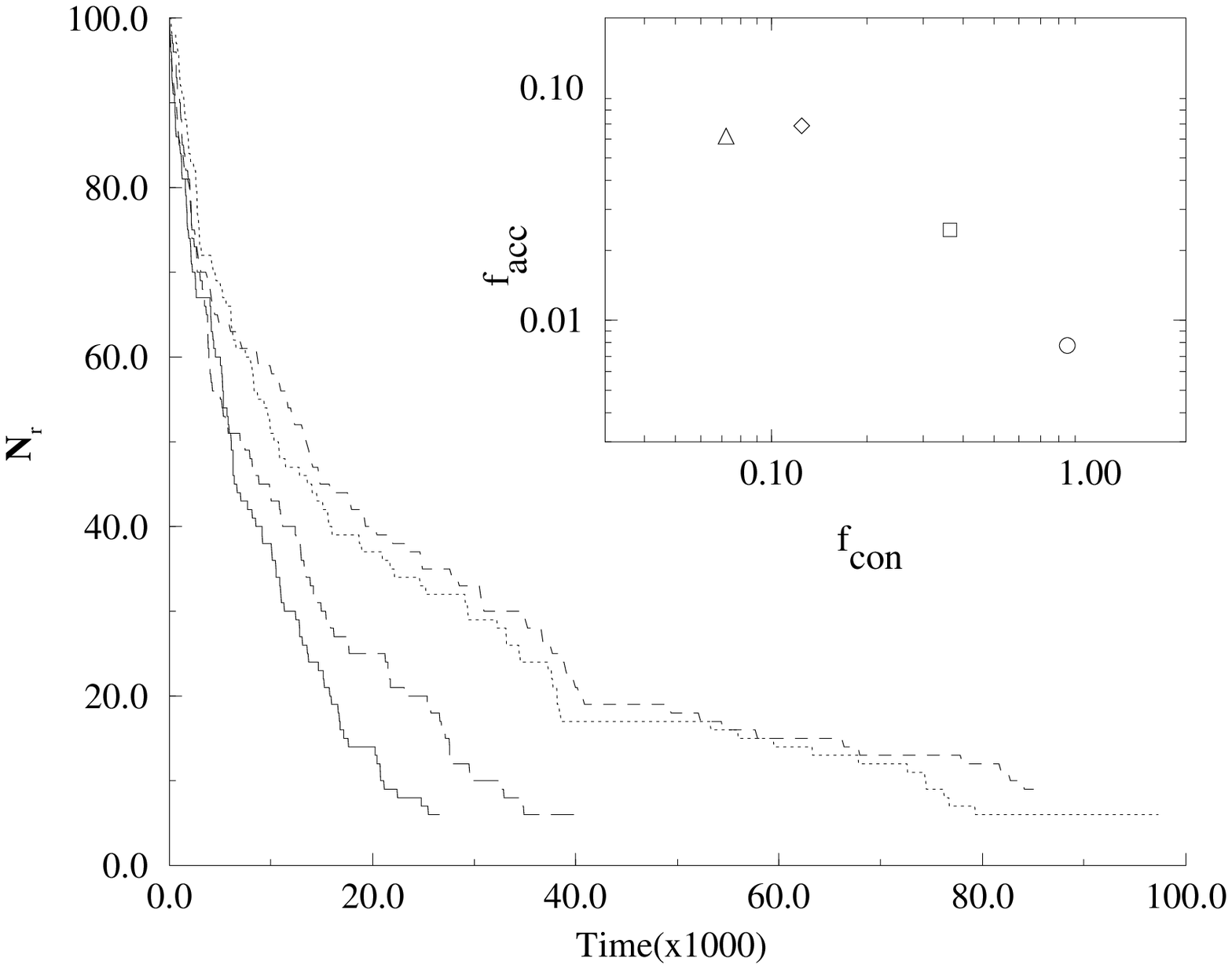,width=.9\linewidth}
\end{center}
\end{figure}

\begin{figure}[htbp]
\begin{center}
  \epsfig{file=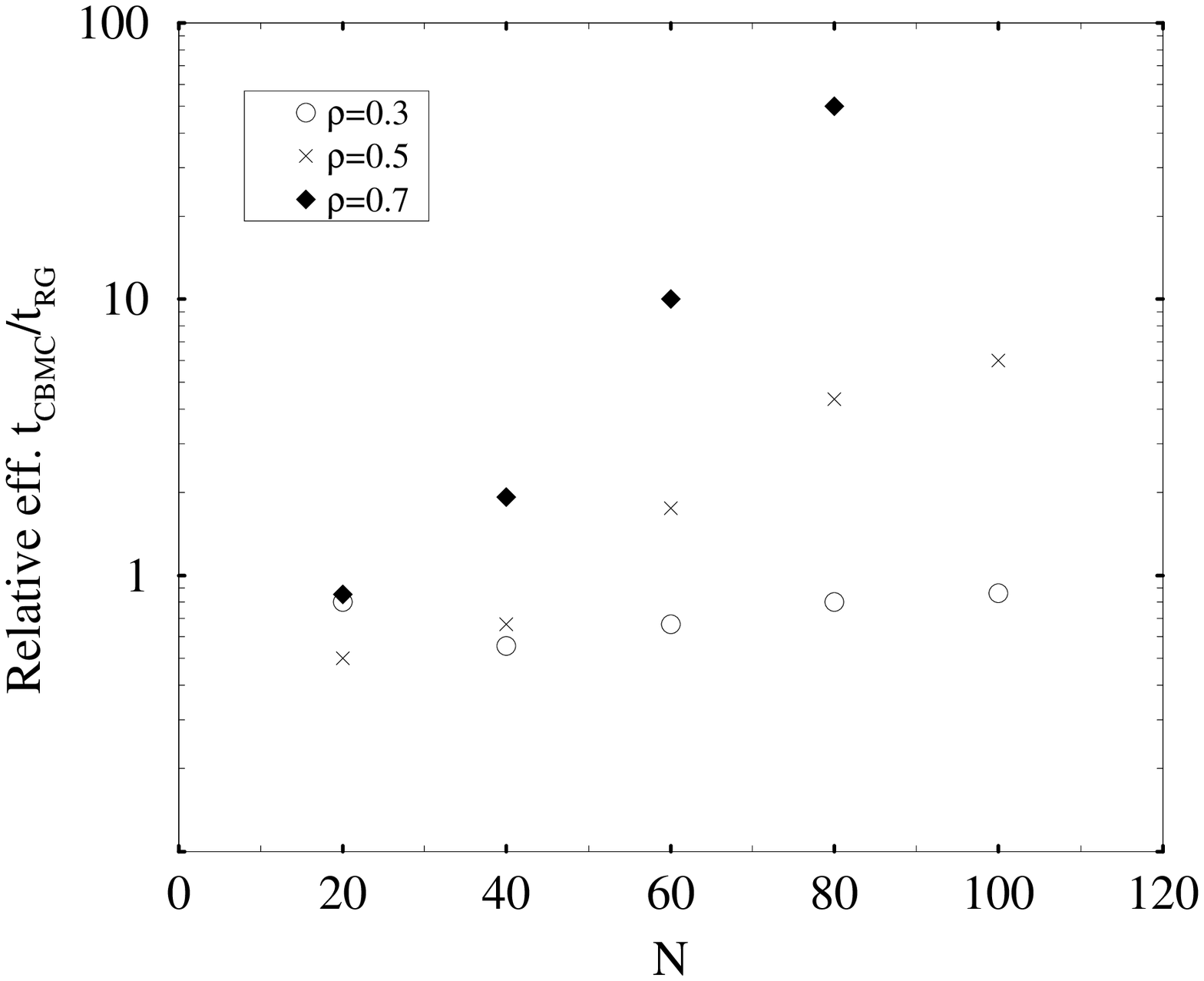,width=.9\linewidth}
\end{center}
\end{figure}

\clearpage
\begin{center}
    \begin{tabular}{|cc|ccccc|}
      \hline\hline
       \multicolumn{2}{|r|}{L}& 20 &   40   &    60    &    80    &   100  \\
       \multicolumn{2}{|l|}{$\rho$}& & & & & \\
      \hline
      $0.3$ & & (2.5,3)&  (2.5,3) & (2.5,4) & (2.3,6)&  (2.4,6) \\
      $0.5$ & & ( 2.7,3)&  (2.5,5) & (2.7,6)& (2.7,9) &   (2.7,10) \\
      $0.7$ & & (3.1,4) & (3.9, 4) & (3.75, 6) & (3.5, 8) & (4.9,10) \\
      \hline\hline
    \end{tabular}
\end{center}


\begin{thebibliography}{99}
  
\bibitem{BOOK} {\em Monte Carlo and Molecular Dynamics simulations in
    Polymer Science}, ed. K. Binder, (Oxford University Press, Oxford, 1995).
  
\bibitem{KREMER} K. Kremer and K. Binder, Comp. Phys. Rep. {\bf 7},
  259 (1988).
  
\bibitem{LEONTIDIS} E. Leontidis, B.M. Forrest, A.H. Widmann and U.W.
  Suter, J. Chem. Soc. (Faraday Transactions) {\bf 91}, 2355 (1995).
  
\bibitem{LAL} M. Lal and R.F.T. Stepto, J. Polym. Sci.: Polym. Symp.
  {\bf 61}, 401 (1977).
  
\bibitem{DEUTSCH} J.M. Deutsch, J. Phys. (Paris) {\bf 48}, 141 (1987).
  
\bibitem{SOKAL} A.D. Sokal in \protect\cite{BOOK}
  
\bibitem{PANT} P.V. Krishna Pant and D. N. Theodorou, Macromolecules,
  {\bf 28}, 7224 (1995).
  
\bibitem{MANSFIELD} M. L. Mansfield, J. Chem. Phys. {\bf 77} 1554
  (1982).
  
\bibitem{PAKULA} T. Pakula, Polymer {\bf 28} 1293 (1987).

  
\bibitem{ROSENBLUTH} Rosenbluth, M.N, and A.W. Rosenbluth, J. Chem.
  Phys. {\bf 23}, 356 (1955).
  
\bibitem{MEIROVITCH} H. Meirovitch, J. Phys. {\bf A15}, L735 (1982).
  
\bibitem{DSM} H. Meirovitch, J. Chem. Phys. {\bf 89}, 2514 (1988).

\bibitem{SIEPMANN} J. I. Siepmann, Mol. Phys. {\bf 70}, 1145 (1990).
  
\bibitem{SIEPMANN92} J. I. Siepmann, D. Frenkel, Mol. Phys. {\bf 75},
  59 (1992).
  
\bibitem{FRENKEL92} D. Frenkel and B. Smit, Mol. Phys. {\bf 75}, 983
  (1992)
  
\bibitem{FRENKEL91} D. Frenkel, G.C.A.M. Mooij, and B. Smit, J. Phys:
  Condens. Matter {\bf 4}, 3053 (1991).
  
\bibitem{BATOULIS} J. Batoulis and K. Kremer, J. Phys. {\bf A21}, 127
  (1988).
  
  
\bibitem{DIJKSTRA} M. Dijkstra, D. Frenkel, and J.-P. Hansen, J. Chem.
  Phys. {\bf 101}.  3179 (1994).
  
\bibitem{VENDRUSCOLO} M. Vendruscolo, J. Chem. Phys., {\bf 106}, 2970
  (1997).
  
\bibitem{ALEXANDROWICZ} Z. Alexandrowicz, J. Chem. Phys., {\bf 109}, 5622 (1998).
  
\bibitem{HEMMER} S. Hemmer and P.C. Hemmer, J. Chem. Phys. {\bf 81}
  585 (1984); P.C. Hemmer and S. Hemmer, Phys. Rev. {\bf A34}, 3304
  (1986).

  
\bibitem{DAANSBOOK} D. Frenkel and B. Smit, {\em Understanding
    Molecular Simulation}, Academic Press, Boston, (1996).

  
\bibitem{SHELLEY}J.C. Shelley and G.N. Patey, J. Chem. Phys., {\bf
    102}, 7656, (1995).

  
\bibitem{SMIT} B. Smit, Mol. Phys. {\bf 85}, 153 (1995).

  
\bibitem{MOOIJ1} G.C.A.M. Mooij, D. Frenkel and B. Smit, J. Phys.
  Condens. Matter {\bf 4}, L255 (1992).
  
\bibitem{WALL} F.T. Wall and J.J. Erpenbeck, J. Chem. Phys. {\bf 30},
  634 (1959).
  

\bibitem{GRASSBERGER} P. Grassberger, Phys. Rev. {\bf E56}, 3682
  (1997); P. Grassberger, H. Frauenkron and W. Nadler,
  cond-mat/9806321.
  
\bibitem{WILDING} N.B. Wilding, M. M\"{u}ller and K. Binder, J. Chem.
  Phys. {\bf 105}, 802 (1996).

\bibitem{MOOIJ96} G.C.A.M. Mooij and D. Frenkel, Mol. Simul., {\bf 17},
41, (1996).





  

\end{thebibliography}
\end{document}